\documentclass[preprint]{aastex}

\usepackage{epsf}

\def\linebreak{\hfil\break}

\def\
{\hfil\linebreak}

\def\degree{\ifmmode {^\circ}\else {$^\circ$}\fi}
\def\mum{\ifmmode {\rm \mu {\rm m}}\else $\rm \mu {\rm m}$\fi}
\def\arcsec{\ifmmode ^{\prime \prime}\else $^{\prime \prime}$\fi}

\def\inch{\ifmmode ^{\prime \prime}\else $^{\prime \prime}$\fi}
\def\arcmin{\ifmmode ^{\prime}\else $^{\prime}$\fi}

\def\msun{\ifmmode {\rm M_{\odot}}\else $\rm M_{\odot}$\fi}
\newbox\grsign \setbox\grsign=\hbox{$>$} \newdimen\grdimen \grdimen=\ht\grsign
\newbox\simlessbox \newbox\simgreatbox
\setbox\simgreatbox=\hbox{\raise.5ex\hbox{$>$}\llap
     {\lower.5ex\hbox{$\sim$}}}\ht1=\grdimen\dp1=0pt
\setbox\simlessbox=\hbox{\raise.5ex\hbox{$<$}\llap
     {\lower.5ex\hbox{$\sim$}}}\ht2=\grdimen\dp2=0pt

\def\simless{\mathrel{\copy\simlessbox}}

\begin{document}

\title{Collisional Cascades in Planetesimal Disks 
II. Embedded Planets}
\vskip 7ex
\author{Scott J. Kenyon}
\affil{Smithsonian Astrophysical Observatory,
60 Garden Street, Cambridge, MA 02138} 
\email{e-mail: skenyon@cfa.harvard.edu}

\author{Benjamin C. Bromley}
\affil{Department of Physics, University of Utah, 
201 JFB, Salt Lake City, UT 84112} 
\email{e-mail: bromley@physics.utah.edu}
%
\author{}
\affil{Astronomical Journal, January 2004}
%
%

\begin{abstract}

We use a multiannulus planetesimal accretion code to 
investigate the growth of icy planets in the outer regions 
of a planetesimal disk. In a quiescent minimum mass solar nebula,
icy planets grow to sizes of 1000--3000 km on a timescale
$t_P \approx 15-20 ~ {\rm Myr} ~ ( a / {\rm 30 ~ AU})^3$,
where $a$ is the distance from the central star. Planets form
faster in more massive nebulae.  Newly-formed planets stir up
leftover planetesimals along their orbits and produce a
collisional cascade where icy planetesimals are slowly ground
to dust. 

The dusty debris of planet formation has physical characteristics
similar to those observed in $\beta$ Pic, HR 4796A, and other
debris disks.  The computed dust masses are
$M_d (r \lesssim {\rm 1 ~ mm}) \sim 10^{26}$ g $(M_0 / M_{MMSN})$ and
$M_d ({\rm 1 ~ mm} \lesssim r \lesssim {\rm 1 ~ m}) \sim 10^{27}$ g 
$(M_0 / M_{MMSN})$, where $r$ is the radius of a particle, 
$M_0$ is the initial mass in solids, and $M_{MMSN}$ is the mass
in solids of a minimum mass solar nebula at 30--150 AU.
The luminosity of the dusty disk relative to the stellar luminosity
is $L_D/L_0 \sim L_{max} (t / t_0)^{-m}$,
where $L_{max} \sim 10^{-3} (M_0 / M_{MMSN})$,
$t_0 \approx$ 10 Myr to 1 Gyr, and $m \approx$ 1--2.
Our calculations produce bright rings and dark gaps with
sizes $\Delta a/a \approx$ 0.1. Bright rings occur
where 1000 km and larger planets have recently formed.
Dark gaps are regions where planets have cleared out
dust or shadows where planets have yet to form.

Planets can also grow in a planetesimal disk perturbed by the close 
passage of a star.  Stellar flybys initiate collisional cascades, 
which produce copious amounts of dust. The dust luminosity following
a modest perturbation is 3--4 times larger than the maximum
dust luminosity of a quiescent planet-forming disk.  In 10 Myr 
or less, large perturbations remove almost all of the planetesimals 
from a disk. After a modest flyby, collisional damping reduces 
planetesimal velocities and allows planets to grow from the 
remaining planetesimals.  Planet formation timescales are 
then 2--4 times longer than timescales for undisturbed disks; 
dust luminosities are 2--4 times smaller.

\end{abstract}

\subjectheadings{planetary systems -- solar system: formation -- 
stars: formation -- circumstellar matter}

\section{INTRODUCTION}

Observations demonstrate that most stars are born
with circumstellar disks of gas and dust.
The disks have masses and sizes similar to the protosolar 
nebula that formed our solar system \citep{bec99,lad99}.  
As they age, stars lose their circumstellar disks.  In 
nearby young star clusters, 1 Myr old stars often have 
massive optically thick disks; 10 Myr old stars rarely 
have opaque disks \citep{hai01}.  
However, older stars often have a `debris disk' 
of dusty material with a size comparable to the
radius of the Kuiper Belt of our solar system
\citep{hab01,so00,spa01,luu02}.  
Debris disk masses are at least a few lunar masses and
may exceed the mass, $\sim$ 0.1 $M_{\oplus}$, of the 
Kuiper Belt \citep{bac93,art97,lag00,luu02,wya03a,gre03}.

Recent imaging and photometric data suggest that debris
disks often have internal structure.  Several disks have 
asymmetries and warps in their surface brightness
distributions \citep{lag00,aug01}.  Distinct rings of dust 
surround $\beta$ Pic \citep{kal00,wah03},
$\epsilon$ Eri \citep{gre98},
HR 4796A \citep{jay98,koe98,au99,sch99,gre00}, 
and Vega \citep{wil02,koe01}.
HD 141569 has a dark gap in its disk \citep{wei99}.
Photometric eclipses in KH 15D suggest at least one
discrete, opaque object in orbit around the central 
pre-main sequence star \citep{her02}.  The nearby A-type
star Fomalhaut also has a distinct clump in its debris disk
\citep{wya02,hol03}.

Producing internal structure in a dusty debris disk requires
a two (or more) step process.  Because radiation removes dust 
from the disk on timescales shorter than the stellar age, 
some process must replenish the dust.  Collisions between large
bodies form copious amounts of dust if the collision velocity 
is $\sim$ 100--300 m s$^{-1}$ \citep[e.g.][]{tan96,ken99,kri00}.
Small embedded planets \citep{ken99} and close encounters with
field stars \citep{ida00,kal00} can excite large collision 
velocities and initiate a `collisional cascade,' where 1--10 km
planetesimals are slowly ground into fine dust grains. The 
reservoir of mass needed to maintain a collisional cascade for 
the 100--500 Myr lifetimes of most known debris disk systems 
is $\sim$ 10--100 $M_{\oplus}$ \citep{hab01,spa01,wya03a}. 
This mass is $\sim$ 10\% to 100\% of the amount of solid material 
in the `minimum mass solar nebula', the minimum amount of 
material needed to form the planets in our solar system scaled 
to solar abundances \citep{wei77a,hay81}.

During the collisional cascade, dynamical processes impose
structure on the dust.  The gravity of a passing star can
produce ring-like density concentrations in a particle
disk \citep{kal00}. Radiation pressure ejects small dust
grains and forms dusty rings at heliocentric distances of 
80--100 AU \citep{tak01}. Small planets stir up leftover
planetesimals along their orbits and produce bright rings 
of dust \citep{kb02b}.  Larger planets or shepherd moons
can produce dark gaps, partial rings, or warps in the disk
\citep{wil02,oz00,wya99,wya03b}. 

To investigate formation mechanisms for debris disks, we consider 
dusty disks as plausible remnants of recent planet formation.
In the planetesimal theory, planets 
grow from mergers of smaller objects embedded in the disk 
\citep{saf69}.  Small dust grains in the disk grow to mm sizes 
and settle into a thin, dusty layer at the midplane of the disk.  
Collisions between grains in the midplane may form successively 
larger grains which grow into 1~km `planetesimals' 
\citep{wei80,wei93,kor01}. Planetesimals may form directly through 
gravitational instabilities \citep{gol73,you03}.  In either case, 
slowly-moving planetesimals collide and merge to form larger 
and larger bodies which eventually become planets 
\citep{wet93,wei97,kl99}.
The gravity of 1000 km or larger planets stirs up leftover 
planetesimals to large velocities \citep{kb01}.  Collisions 
between rapidly moving planetesimals produce observable amounts 
of dust, which often lie in ring-like structures \citep{ken99,kb02b}. 

In this paper, we continue our study of the bright rings and 
dark gaps formed during the growth of icy planets in the
outer regions of a planetesimal disk. \cite{kb02b} show that
icy planets reach sizes of 1000--3000 km on timescales of 
$\sim$ 10--20 Myr at 30 AU and 500 Myr to 1 Gyr at 100 AU. 
The dusty debris of icy planet formation often lies in bright
rings with lifetimes of 10--50 Myr. Dark gaps between the bright 
rings indicate an absence of dust.  Here we show that bright, 
optically thick rings can shadow the disk and produce apparent 
dark gaps in the disk.  The rings and gaps produced in our models 
are comparable in size and surface brightness to those observed 
in some debris disks, such as HR 4796A.  We also derive predicted
lifetimes, luminosities, and masses for collisional cascades induced 
by embedded planets or a stellar flyby. Although our derived lifetimes
are a factor of $\sim$ 2 longer than observed, our results for 
dust luminosities and masses agree with observations 
\citep[e.g.][]{gre03,hab01,wya03a,dec03}. 

We outline the model in \S2, describe the calculations in \S3,
and conclude with a brief discussion in \S4.  

\section{THE MODEL}

\citet[][2002a]{kb01} describe our multiannulus numerical model for 
planetesimal growth\footnote{\citet{the03} describe a single
annulus model in the context of the $\beta$ Pic disk.}.  
\citet[][1999]{kl98}, \citet{ken02}, and \citet[][2002a]{kb01}
compare results with analytical and numerical calculations.
Briefly, we adopt the \citet{saf69} statistical approach to calculate 
the collisional evolution of an ensemble of planetesimals in orbit
around a star of mass $M$ \citep[see also][]{spa91,wei97}. The
model grid contains $N$ concentric annuli with widths $\delta a_i$
centered at heliocentric distances $a_i$.  Calculations begin with 
a differential mass distribution $n(m_{ik}$) of bodies with horizontal 
and vertical velocities $h_{ik}(t)$ and $v_{ik}(t)$ relative to a circular
orbit.  The horizontal velocity is related to the orbital eccentricity,
$e_{ik}^2(t)$ = 1.6 $(h_{ik}(t)/V_{K,i})^2$, where
$V_{K,i}$ is the circular orbital velocity in annulus $i$.
The orbital inclination depends on the vertical velocity,
$i_{ik}^2(t)$ = sin$^{-1}(2(v_{ik}(t)/V_{K,i})^2)$.

The mass and velocity distributions evolve in time due to inelastic 
collisions, drag forces, and long-range gravitational forces.  We derive 
collision rates from kinetic theory and use an energy-scaling algorithm 
to choose among possible collision outcomes.  We define $S_0$ as the
tensile strength of a planetesimal and $E_g$ as the gravitational
binding energy per unit mass. If $E_c$ is the center of mass collision
energy, the ratio $x_c = E_c/(E_g+S_0)$ sets the collision outcome;
$x_c \ll$ 1 yields a merger with negligible debris,
$x_c \sim$ 1 yields a merger with some debris, and
$x_c \gg$ 1 yields only debris.  The ratio of the collision energy
$Q_f$ to the crushing energy $Q_c$ defines the mass of the debris,
$m_d = Q_f/Q_c$.  For most calculations, we adopt a crushing energy 
$Q_c = 5 \times 10^7$ erg g$^{-1}$ and a range of $S_0$ appropriate 
for icy objects at large distances from the central star,
$S_0 \sim$ 1 to $10^6$ erg g$^{-1}$ \citep{gre84,kl99}.  
Most fragmentation 
algorithms adopt a minimum velocity for fragmentation, $V_f$
\citep{dav85,wet93}; we choose $V_f$ = 1 cm s$^{-1}$
\citep{kl99}.

We use two algorithms to compute the velocity evolution from long-range
gravitational interactions.  In the high velocity regime, the collision
velocity exceeds the mutual Hill velocity,
\begin{equation}
v_H \approx \Omega_{ij} a_{ij} [(m_{ik} + m_{jl})/3 M_{\oplus}]^{1/3} ~ ,
\end{equation}
where $\Omega_{ij}$ are $a_{ij}$ are the average angular velocity and
the average heliocentric distance of annulus $i$ and annulus $j$.  
Statistical solutions to the Fokker-Planck equation
then yield accurate estimates for the stirring rates \citep[e.g.,][]{ste00}.
In the low velocity regime, the Fokker-Planck equation underestimates
the stirring timescales.  For some of our calculations, we adopt the 
\citet{ida93} fits to low velocity stirring rates
derived from $n$-body simulations.  \citet{kb01} describe how we match
stirring rates in between the low and high velocity regimes. For other 
calculations, we adopt the \citet{oht02} stirring rates derived from
fits to the \citet{ste00} rates and $n$-body calculations.  In most
situations, the \citet{oht02} rates are 20\% to 30\% smaller than the
\citet{kb01} rates. The \citet{oht02} rates also agree better with 
the analytic formulae of \cite{gol02} than the \cite{kb01} rates.

Gas drag circularizes the orbits of all mass batches and also removes 
material from each batch.  We adopt a simple nebular model with
gas surface density $\Sigma_g(a,t) = \Sigma_{g0} a^{-3/2} e^{-t/t_g}$ 
and scale height $H_g(a) = H_0 (a/a_0)^{1.125}$ 
\citep{kh87} to compute the 
gas volume density $\rho_g$. To approximate gas removal on a
time $t_g$, the gas density declines exponentially with time.
We set $t_g$ = 10 Myr and adopt the \citet{ada76} formalism 
to compute inward drift and velocity damping from gas drag
\citep[see][for another discussion of gas drag]{wei77b}.  
In calculations at 30--150 AU, particle losses from gas drag 
are small, $\sim$ 1\% or less of the initial mass. Velocity
damping is negligible at late stages when viscous stirring 
dominates the velocity evolution of small bodies.

The initial conditions for these calculations are appropriate for a
disk with an age of $\sim$ 1--10 Myr.  We consider systems of $N$ annuli 
in disks with $a_i$ = 30--150 AU and $\delta a_i/a_i$ = 0.01--0.025.  
The disk is composed of small planetesimals with radii of $\sim$ 
1--1000 m (see below).  The particles have an initial mass distribution 
$n_i(m_{ik})$ in each annulus, with a mass ratio $\delta = m_{ik+1}/m_{ik}$
between adjacent bins. These objects begin with eccentricity $e_0$ and 
inclination 
$i_0 = e_0/2$.  Most of our models have $e_0$ independent of $a_i$; some 
models have a constant initial horizontal velocity in each annulus,
$e_i \propto a_i^{1/2}$.
We assume a power law variation of the initial surface density of 
solid material with heliocentric distance, $\Sigma_i = \Sigma_0$ 
$(a_i/a_0)^{-3/2}$.  Models with $\Sigma_0 \approx$ 0.1--0.2 g cm$^{-2}$
at $a_0$ = 30 AU have a mass in icy solids comparable to the
minimum mass solar nebula \citep{wei77a,hay81}.  Observed disk masses
for 1--10 Myr old stars range from $\sim$ 10\% to $\sim$ 1000\% of the 
minimum mass solar nebula \citep{ost95,nat97,wya03a}.

Our calculations follow the time evolution of the mass and velocity 
distributions of objects with a range of radii, $r_{ik} = r_{min}$
to $r_{ik} = r_{max}$.  The upper limit $r_{max}$ is always larger
than the largest object in each annulus.  To save computer time in
our main calculation, we 
do not consider small objects which do not affect significantly the
dynamics and growth of larger objects, $r_{min}$ = 10--100 cm.  
Erosive collisions produce objects with $r_{ik}$ $< r_{min}$ which
are `lost' to the model grid. Lost objects are more likely to be 
ground down into smaller objects than to collide with larger objects 
in the grid.

To estimate the amount of dusty debris produced by planet formation,
we perform a second calculation. Each main calculation yields
$\dot{M}_i (t)$, the amount of mass lost to the grid per annulus 
per timestep.  The total amount of mass lost from the planetesimal
grid is $\dot{M} = \sum_{i=1}^{N} \dot{M}_i (t)$.  The debris has 
a known size distribution, 
$n^{\prime}_{ij} = n^{\prime}_{i0} a_i^{-\beta}$, where
$\beta$ is a constant.  The normalization constant $n_{i0}^{\prime}$
depends only on $\beta$ and $\dot{M} (t)$, which we derive at
each timestep in the main calculation.  To evolve the dust 
distribution in time, we use a simple collision algorithm. 
The optical depth $\tau$ of the dust follows from integrals over 
the size distribution in each annulus. The optical depth and a
radiative transfer solution then yield the luminosity and
radial surface brightness of the dust as a function of time.  
The appendix describes the collision algorithm and the 
calculations for the optical depth and the dust luminosity.

\section{CALCULATIONS}

\subsection{Planet Formation in a Quiescent Disk}

\subsubsection{Wetherill \& Stewart Fragmentation}

We begin with calculations in a quiescent disk
surrounding a star with a mass of 3 $M_{\odot}$ and a
luminosity of 50 $L_{\odot}$.
The disk consists of 64 annuli with $\Delta a_i/a_i$ = 0.025
and extends from 30 AU to 150 AU.  The initial distribution of 
icy planetesimals has sizes of 0.5--1000 m with $\delta = 2$ 
and equal mass per mass bin. The initial eccentricity is
$e_0 = 10^{-5}$ for all objects in all annuli; the initial 
inclination is $i_0 = e_0/2$. These initial values provide
a rough equilibrium between collisional damping and viscous
stirring at $t = 0$. Viscous stirring dominates at smaller
$e_0$; collisional damping dominates at larger $e_0$.
The particles have mass density
$\rho_d$ = 1.5 g cm$^{-3}$. The initial surface density in solid
objects is comparable to the minimum mass solar nebula, with
$\Sigma_0$ = 0.18 g cm$^{-2}$ ($a_0$/30 AU)$^{-3/2}$.  The total mass 
in solids, 93.8 $M_{\oplus}$, is similar to the median dust mass 
observed in disks around young 0.5--3 $M_{\odot}$ stars \citep{wya03a}.
We adopt the \cite{ada76} gas drag formalism, the \cite{oht02} 
stirring rates, and the \citet{wet93} fragmentation algorithm 
with $Q_c = 5 \times 10^7$ erg g$^{-1}$ and 
$S_0 = 10^6$ erg g$^{-1}$.  These standard choices for $Q_c$ 
and $S_0$ are appropriate for icy planetesimals with bulk 
properties similar to terrestrial snow \citep{gre84}.

Icy planet formation in the outer regions of a quiescent disk
has three stages \citep{kl99,ken02}. Small planetesimals with
$r_i \approx$ 1--1000 m and $e_0 \lesssim 10^{-3}$ grow slowly.
At 30--37 AU, it takes $\sim$ 0.3 Myr for the largest objects
to grow from $r_i \sim$ 1 km to $r_i \sim$ 10 km. Slow, orderly
growth ends when the gravitational cross-sections of the largest 
objects exceed their geometric cross-sections. Runaway growth
begins.  The largest objects take $\sim$ 2 Myr to reach
sizes of 100 km and another 15 Myr to reach sizes of
1000 km.  As the largest objects grow rapidly, dynamical
friction and viscous stirring increase the eccentricities
of the smallest objects.  Collisions between these small
objects then begin to produce more and more debris. As the
largest objects grow to sizes of 2000--3000 km, a collisional
cascade reduces substantially the mass in the smallest objects.
During this period of `oligarchic growth \citep{kok98},' 
the system reaches a rough equilibrium where the largest 
objects contain most of the remaining mass.

Figure 1 illustrates the time evolution of the cumulative
number and eccentricity distributions at 30--37 AU in the 
inner disk.  The initial distributions are 
$N_C \propto r_i^{-q_0}$ with $q_0$ = 3 and
$e_0 = 10^{-5}$ for all objects.  During slow growth,
the velocity distribution develops two features (Figure 1; 
right panel). For the largest objects,
dynamical friction dominates viscous stirring and produces 
a power law velocity distribution. At smaller radii, viscous
stirring produces a flat velocity distribution with
$e_{flat} / e_{min} \sim$ 0.1 \citep{gol02}. The transition 
between these two regimes occurs at the particle radius 
where most of the mass is concentrated.
During runaway growth, this transition moves to larger 
$r_i$; the shape of the velocity distribution does not change.

The mass distribution consists of two power laws,
with a transition at $r_i \sim$ 1 km (Figure 1; 
left panel). The largest objects always follow 
a power law with $\alpha_l \sim 3$. 
At the start of the slow growth phase, collisions
produce mergers and little debris. The power law exponent
for the small bodies thus decreases from the initial
$\alpha_s \approx$ 3 to $\alpha_s \approx$ 1--2.
During runaway growth, debris from collisions adds mass to 
the smallest mass bins; $\alpha_s$ slowly increases to the 
standard collisional exponent, $\alpha_s \sim$ 2.5 
\citep{doh69,wil94}. Although collisions between small 
objects reduce the total mass contained in small bodies, 
the slope of the mass distribution is roughly constant,
with $\alpha_s \approx 2.5$.

Figure 2 shows the evolution of the cumulative surface
density in the inner disk.  During the first 1 Myr of
evolution, slow growth concentrates most of the mass in 
1--10 km objects. Because most collisions yield mergers 
with little debris, the total surface density is
roughly constant. Runaway growth concentrates more
mass in the largest objects.  After 10--20 Myr, large objects 
with radii of 100--1000 km contain roughly 10\% of the 
initial mass in the inner disk.  Once oligarchic growth
begins, collisions between small objects produce substantial
debris instead of mergers. The surface density begins to
decline. The largest objects continue to grow and contain
an ever-larger fraction of the total mass. After 1--2 Gyr,
disruptive collisions between small objects reduce the 
surface density by nearly an order of magnitude.
Large objects with $r_i \gtrsim$ 100 km contain $\sim$ 10\%
of the initial mass and $\sim$ 70\% of the final mass
of solid material at 30--37 AU.

The timescale for planet growth is a strong function of
heliocentric distance. Because the collision
time is proportional to the surface density and the
orbital period, the growth time scales with $a$ as
$t \propto P / \Sigma $ $\propto a^3$ \citep{lis87,kl98}.  
Planetesimals
thus grow to large sizes first in the inner disk. In this
model, the ratio of timescales for planets to grow to 
1000 km at 135 AU and at 33 AU is $\sim$ 80.  This result 
is close to the predicted ratio of $\sim 4^3$ = 64.

Despite differences in timescales, planet formation
proceeds to the same endpoint in all annuli.  Slow
growth, runaway growth, and oligarchic growth produce
large objects with $r_i \sim$ 100 km to $\sim$ 
3000 km which contain $\sim$ 5\%--10\% of the initial mass.  
These large objects stir up the orbits of the leftover
planetesimals and initiate a collisional cascade.  These 
disruptive collisions reduce the surface density of small 
objects by roughly an order of magnitude at all $a$.

Erosive collisions also lead to structure in the disk 
(Figure 3).  During the slow growth stages, most 
collisions yield mergers.  The production rate $\dot{M}$
of objects with $r_i \lesssim$ 1 m is small and 
declines smoothly with increasing heliocentric distance.  
As planetesimals merge to form planets, the collisional
cascade produces more and more debris.  At the inner edge 
of the disk, $\dot{M}$ increases by more than 3 orders of 
magnitude in $\sim$ 50 Myr.  The collisional cascade
rapidly depletes the inner disk of 1--1000 m bodies;
$\dot{M}$ declines.  As large planets begin to form in 
the outer disk, the peak in $\dot{M}$ moves to larger 
heliocentric radii.  This peak moves from $a_i \sim$ 30 AU
at $t$ = 80 Myr, to $a_i \sim$ 50 AU at 400 Myr, and to
$a_i \sim$ 100 AU at $t$ = 2--3 Gyr.  

\subsubsection{Davis {\it et al.} Fragmentation}

Calculations with the \citet{dav85} fragmentation algorithm 
allow us to measure the sensitivity of the results to the 
bulk properties of  the planetesimals. In the \citet{dav85}
approach, the ejecta from a collision receive a fixed fraction 
$f_{KE}$ of the center-of-mass collision energy.  We follow 
\citet{kl99} and adopt $f_{KE}$ = 0.05 and $S_0$ = 1--$10^6$
erg g$^{-1}$ to simulate a wide range of planetesimal bulk
properties \citep[see also][]{gre84,dav85}.

Planets form faster with the \citet{dav85} fragmentation 
algorithm \citep[Figure 4; see also][]{kl99}. During the 
slow growth phase, collisions that produce some debris 
tend to reduce $e$ and $i$ for the largest planetesimals 
and increase $e$ and $i$
for the smallest planetesimals. Cooling of the large 
planetesimals increases gravitational focusing factors
and speeds the growth of the largest objects during
runaway growth. As a result, it takes only $\sim$ 15 Myr
for planetesimals to grow to sizes of 1000 km with the
\citet{dav85} fragmentation algorithm and $S_0 = 10^6$
erg g$^{-1}$, compared to $\sim$ 20 Myr with the 
\citet{wet93} algorithm.

With both fragmentation 
algorithms, collisional cascades remove nearly all of the mass 
from a planetesimal disk.  Figure 5 compares derived values 
of $\dot{M}$ for models with a range of $S_0$. In all models,
significant debris production begins when large objects
with $r_i \gtrsim$ 100 km stir small objects to the 
disruption velocity, $V_d \approx$ 50 m s$^{-1}$ 
$(S_0 / 10^4 ~ {\rm erg ~ g^{-1}})^{1/2}$ \citep{dav85,kb01}.
Calculations with small $S_0$ thus produce more debris
at earlier times than calculations with large $S_0$. 
Planetesimals are easier to fragment, but harder to
disrupt, in the \citet{dav85} fragmentation algorithm
than in the \citet{wet93} algorithm. At fixed $S_0$,
models with \citet{dav85} fragmentation produce more debris 
at early times, and somewhat less debris at later times,
than \citet{wet93} models.  

Despite these differences, all calculations have several 
common features.  Shortly after the debris production rate 
reaches a maximum, the decline in $\dot{M}$ follows a power
law with $\dot{M} \propto t^{-1}$. At $t \approx$ 100 Myr, 
all models converge to $\dot{M} \sim 5 \times 10^{20}$ g yr$^{-1}$. 
Over a 6 order of magnitude range in $S_0$, the dispersion
in $\dot{M}$ at $t \approx$ 100 Myr is smaller than a factor 
of two.  This common $\dot{M}$ leads to a standard dust
mass and disk luminosity at $t \sim$ 10--100 Myr (see below).

Figure 6 compares final cumulative surface density distributions
for the inner eight annuli of several models.  For $S_0$ = 
$10^6$ erg g$^{-1}$, 1 Gyr of fragmentation reduces the 
surface density of solid material in the disk by roughly 
an order of magnitude. For $S_0$ = 1 erg g$^{-1}$, the 
cumulative surface density is roughly a factor of 30 smaller 
than the initial surface density.  Despite a large range in 
$S_0$, the range in the final cumulative surface density is 
small.  In these models, large bodies with $r_i \gtrsim$ 100 km 
contain $\sim$ 1\% to 10\% of the initial mass and $\sim$ 
30\% to 80\% of the final mass in solid material.

Stochastic processes are an important feature of the
collisional cascade.  During runaway growth,  random 
fluctuations in the collision rate can produce a single 
large body which grows much more rapidly than bodies 
in neighboring annuli. By robbing other bodies of material, 
this runaway body slows the growth of large objects nearby.  
Stirring by the runaway object leads to more dust on shorter 
timescales compared to calculations without a single runaway.  
Single runaways occur in $\sim$ 20\% to 25\% of test 
calculations with identical initial conditions.  These runaways 
appear to occur more often in models with smaller $S_0$.

The general results of the calculations are insensitive to 
changes in the initial conditions.  For initial eccentricity
$e_0 \le 10^{-3}$, all calculations yield growth by mergers
instead of rapid disruption \citep[see][]{kb02a}. In models
with $e_0 \sim 10^{-4}$ to $10^{-3}$, collisional damping
reduces $e$ until runaway growth begins to produce large 
objects. Compared to calculations with $e_0$ = $10^{-5}$,
growth times are factors of 2--3 longer \citep[see also][]{kl99}.
Variations in the initial surface density, the mass density,
or the size distribution of planetesimals change the timescale 
but not the outcome of the evolution. Growth times scale with
the initial radius $r_{i0}$ of the largest object in the grid. 
For calculations with $r_{min}$ = 0.5--1.0 m, models with $r_{i0} 
\sim$ 10 m to 1 km have indistinguishable growth times when the
slope of the initial power law mass distribution is $q_0 \lesssim$
4--5.  Because the timescale for viscous 
stirring is shorter than the growth timescale for $r_i \gtrsim$
10 km, models with a substantial fraction of the initial mass 
in 10 km or larger objects take 2--3 times longer to reach 
runaway growth.  Growth rates are more sensitive to the initial 
mass in solids and the mass density of individual planetesimals, 
$t \propto \Sigma^{-1}$ and $t \propto \rho_g^{-2/3}$
\citep[see][]{kl99}.

The results are also insensitive to the details of the gas drag 
and gravitational stirring algorithms.  Because gas drag removes 
only a few per cent of the initial mass in the disk, changes 
to the damping or drag algorithm do not change the results.  
The stirring algorithm affects timescales for the collisional 
cascade.  Calculations with the \citet{ida93} fits for low 
velocity stirring yield smaller stirring rates than the 
Ohtsuki et al. (2002) rates. Smaller stirring rates delay 
the collisional cascade but yield similar debris production.  
For the range of stirring rates we use in our calculations, 
the delay in the collisional cascade is $\sim$ 25\%.

\subsubsection{Luminosity and Surface Brightness Evolution}

Our coagulation calculations demonstrate that planet formation
in the outer regions of a planetesimal disk produces copious 
amounts of dust.  The formation of 1000--3000 km bodies leads 
to a collisional cascade with $\dot{M} \sim$ $10^{18}$ g yr$^{-1}$ 
to $10^{20}$ g yr$^{-1}$ at 30--100 AU. For dust survival times 
of $\sim$ 1 Myr at 30--100 AU, these $\dot{M}$'s imply dust 
masses of $10^{-3} M_{\oplus}$ to $10^{-1} M_{\oplus}$.  
This range is comparable to dust masses derived from 
observations of debris disks \citep{bac93,lag00,zuc01,wya03a}.  
In our simulations, collisional cascades yield large 
$\dot{M}$'s in narrow ranges of disk radii.
The variation in $\dot{M}$ across the face of the disk 
(Figure 3) suggests a ring of debris that expands in 
radius as the formation of 1000--3000 km objects moves 
to larger disk radii.  The apparent dimensions of the 
ring, $\delta r_i \sim$ 20 AU at $r_i \sim $ 50 AU and 
$\delta r_i \sim $ 30 AU at 
$r_i \sim $ 100 AU, are comparable to the dimensions,
$\delta r/r$ $\sim$ 0.1--0.2, of the rings in HR 4796A
and Vega \citep{jay98,koe98,wil02}. 

To quantify comparisons between our models and observations, 
we calculate the evolution of particle numbers for `dust' with 
$r_i \lesssim$ 1 m. The calculation assumes that all small 
particles have a scale height comparable to the scale height 
of the 1 m objects in the planetesimal calculation and follows
the formation and removal of dust grains by collisions and
radiative processes.  Gas drag is not included.  The evolution
of particle numbers yields the radial optical depth through
the disk, which we use to derive the time evolution of the
radial surface brightness $I$ and the fraction of stellar radiation
$L_D/L_0$ absorbed and scattered by the disk. This approach
complements more detailed calculations of dust in a gaseous
disk \citep[e.g.,][]{tak01}. \citet{tak01} solve the equation of 
motion for grains in a gas disk irradiated by a central star
but do not include dust formation by collisions. We derive
dust production rates from the planetesimal evolution code
but do not consider coupling between gas and dust in the disk.
The appendix describes the details of our calculation.

Figure 7 shows the evolution of dust mass in several planetesimal
disks.  We divide the dust into small grains with 
1 $\mu$m $\lesssim r_i \lesssim$ 1 mm and large grains 
with 1 mm $ \lesssim r_i \lesssim$ 1 m.
The small grains have little mass but produce most of the optical 
depth; the large grains contain most of the mass but have small
optical depth.  At $t$ = 0, all disks have no mass in small or
large grains. During the slow growth phase, collisions produce
modest amounts of debris. The dust mass grows linearly in time. 
Runaway growth concentrates more and more
mass into large objects with $r_i \gtrsim$ 100 km, which stir 
the leftover planetesimals to the disruption velocity. Once 
runaway growth begins in the innermost disk annuli, it takes 
only 3--10 Myr for the dust mass to grow by 4--6 orders of
magnitude.  The rapid rise in dust mass begins with the 
formation of objects with $r_i \gtrsim$ 100 km.
When the largest objects have $r_i \sim$ 1000 km at $a_i \sim$ 
30 AU, the dust mass reaches a plateau.  The dust mass is roughly 
constant until large objects start to form at the outer edge 
of the disk.  Poynting-Robertson drag and radiation pressure 
then rapidly remove dust from the disk.  At late times, the 
decline in dust mass is $M_d \propto t^{-n}$, with $n \approx$ 1--2.

The mass in small or large grains is remarkably independent
of the bulk properties of the planetesimals or the initial
conditions in the planetesimal disk.  As planet formation 
propagates through the disk, 
the dust mass is roughly constant in time. Because planetesimals 
with small $S_0$ are easier to fragment than planetesimals 
with large $S_0$, models with $S_0$ = 1 erg g$^{-1}$ reach
this plateau more rapidly ($\sim$ 1 Myr) than models with
$S_0$ = $10^6$ erg g$^{-1}$ (30--50 Myr).  Despite a six order 
of magnitude range in $S_0$, the range in dust mass is a factor 
of two for similar starting conditions.  Minimum mass solar 
nebula models with an initial mass of $\sim$ 100 $M_{\oplus}$ 
in solids at 30--150 AU yield a mass in small grains of 
$M_s \sim$ 0.01--0.02 $M_{\oplus}$; the mass in large grains 
is $M_l \sim$ 0.1--0.25 $M_{\oplus}$.  The dust mass is equally
insensitive to large ranges in $e_0$ or $\rho_d$. The dust mass 
is roughly proportional to the initial mass of solids in the disk,
$M_s,M_l \propto M_0$.  

The dust masses derived from the planetesimal model are
comparable to those observed in debris disk systems.
The masses in small grains are close to the minimum
mass needed to produce an observable mid-IR or far-IR
excess of radiation for a nearby main sequence star
\cite[e.g.,][]{woo02}. 
We derive $M_s \sim$ 0.01--0.02 $M_{\oplus}$ in the 
plateau and $M_s \gtrsim$ $10^{-5} M_{\oplus} $ at late 
stages of the evolution.  \citet{bac93} quote minimum masses 
of $\sim$ 0.001--0.01 $M_{\oplus}$ for $\alpha$ Lyr, 
$\alpha$ PsA, and $\beta$ Pic.  For a sample of remnant 
disks around nearby main sequence stars, \citet{hab01} quote
minimum dust masses of $10^{-4}$--$10^{-2}$ $M_{\oplus}$
\citep[see also][]{gre03,wya03a,dec03}.

During the collisional cascade, large dust masses result in 
luminous disks.  Figure 8 illustrates the time evolution of the
relative disk luminosity $L_D/L_0$ for models with different
$S_0$. The left panel shows luminosity evolution for models
without radiation pressure on small grains. The right panel
shows the luminosity evolution when radiation pressure removes
small grains with $r_i \lesssim$ 1 $\mu$m on the local dynamical
timescale.  Independent of $S_0$, models with radiation pressure
yield maximum disk luminosities of $L_D/L_0 \sim 10^{-3}$. 
Because radiation pressure is unimportant for large $S_0$,
the maximum disk luminosity is independent of radiation forces
for $S_0 \gtrsim 10^4$ erg g$^{-1}$. In models with small $S_0$,
very small grains ejected by radiation pressure contribute most of
the optical depth in the disk. Thus, models without radiation
pressure yield lower disk luminosities for $S_0 \lesssim$
$10^3$ erg g$^{-1}$.

Although the magnitude of the disk luminosity is sensitive to 
the treatment of radiation pressure, the form of the luminosity
evolution is independent of $S_0$ and other parameters in the
calculations.  All models have a relatively rapid rise in
$L_D/L_0$ followed by a longer decline.  While the mass in dust 
is relatively constant in time, the luminosity follows a
shallow power law decline, $L_D/L_0 \propto t^{-m}$, with
$m \approx$ 1.  As planet formation propagates out through the
planetesimal disk, dust forms at larger and larger distances from 
the central star. Because dust in the outer disk intercepts 
less radiation from the central star than dust in the inner 
disk, the dust luminosity declines with time.  
Once Poynting-Robertson drag removes material from the disk,
the luminosity declines more rapidly with a power law index
closer to $m$ = 2.

Collisional cascades also yield a standard maximum luminosity 
for the outer part of the disk. The luminosity is insensitive
to the bulk properties of planetesimals, 
$f_{KE}$, $\rho_d$, $Q_c$, $S_0$, and $V_f$,
and many of the initial conditions,
$e_0$, $q_0$, and $r_{i0}$. The luminosity is sensitive
to the initial mass in planetesimals $M_0$ at 30--150 AU:
\begin{equation}
\frac{L_D (max)}{L_0} \approx 10^{-3} \left ( \frac{M_0}{100 ~ M_{\oplus}} \right )^{-1} ~ .
\end{equation}
This luminosity is comparable to the maximum luminosity of 
known debris disks, such as $\beta$ Pic and HR 4796A.
Because the timescale to reach the maximum luminosity
depends on the initial mass in planetesimals, we can write
the dust luminosity as a function of time,
\begin{equation}
\frac{L_D}{L_0} (t > t_0) \approx 10^{-3} \left ( \frac{t_0}{t} \right )^{-1} ~ ,
\end{equation}
where $t_0 \approx$ 30 Myr $(M_0 / 100 ~ M_{\oplus})^{-1}$.
The coefficient for $t_0$ ranges from 1--3 Myr for 
$S_0$ = 1 erg g$^{-1}$ to $\sim$ 100 Myr for 
$S_0$ = $10^6$ erg g$^{-1}$.  \citet{dom03} derive
similar relations from an analytical model for the
collisional cascade.

Finally, all planetesimal calculations produce axisymmetric
structures within the disk.  Figure 9 shows two sets of 
relative surface brightness distributions for models with different
$S_0$. For $S_0$ = $10^4$ erg g$^{-1}$ (top panel), the
initial surface brightness profile in curve (a) is a power
law, $I/I_0 \propto a_i^{-7/2}$. During the slow growth phase,
the surface area per unit mass of the planetesimals drops 
and the surface brightness fades. Because planets grow faster
in the inner disk, the surface brightness of the inner disk 
fades more rapidly than the brightness in the outer disk. 
The surface brightness reaches a minimum at $t$ = 4 Myr;
$I/I_0 \propto a_i^{-7/3}$.  Once runaway growth begins, 
it takes only 7 Myr for the surface brightness at the inner
edge of the disk to reach a maximum roughly two orders of 
magnitude brighter than the initial planetesimal disk.  
As planet formation propagates through the disk, a bright 
ring of dust emission moves outward. This ring highlights the 
region of maximum dust production and signals the presence
of at least one planet with a radius of 1000 km or larger.
It takes $\sim$ 
60 Myr for this ring to reach $a_i \sim$ 50 AU and another
330 Myr to reach 80 AU. If we define the width $\delta a/a$
of the ring by the radius where the surface brightness drops
to half of the maximum, $\delta a_i/a_i$ $\approx$ 0.2 at $a_i$
= 40 AU and $\delta a_i/a_i$ $\approx$ 0.15 at $a_i$ = 80 AU.
At $t$ = 400 Myr, the inner disk is nearly two orders of 
magnitude fainter than at $t$ = 0.  After 2 Gyr, planet 
formation reaches the outer edge of the disk and the 
surface brightness of the entire disk fades.  As the disk
fades, the surface brightness rises with heliocentric distance,
with $I/I_0 \propto a_i^{p}$ and $p \approx$ 0--2.

Calculations with smaller $S_0$ produce shadowed disks
(Figure 9; bottom panel).
During the early stages of models with $S_0 \lesssim 10^2$
erg g$^{-1}$, the disk surface brightness is a power law and
fades slowly with time (curves (a) and (b)). Once runaway 
growth begins, the inner disk brightens by 2--3 orders of 
magnitude. The dust in the inner disk is optically thick,
$\tau \approx$ 3--10 at $a_i \approx 30$ AU, and shadows 
material in the outer disk.  Shadowing produces a pronounced 
minimum in the surface brightness, which propagates outwards
as planets form at progressively larger disk radii. Although
this shadow resembles the dark gaps cleared of material by a 
large planet, it is not an absence of dust. The shadow is a
region where light from the central star does not penetrate
and is therefore of much lower surface brightness than
surrounding bright regions with comparable amounts of dust.
In models
with $S_0 = 10^2$ erg g$^{-1}$, it takes only $\sim$ 2 Myr 
to form a bright inner ring and a dark shadow at $a_i \approx$
30--40 AU. This structure moves to $a_i \approx$ 50--60 AU 
at 20 Myr and to $a_i \approx$ 80--100 AU at 100 Myr. 
The radial extent of the dark gap is $\delta a/a_i \approx$
0.1.  After 500 Myr to 1 Gyr, planet formation in the outer 
disk starts to produce dust and the entire disk begins to fade.

Figures 10--11 show a montage of color snapshots of the disk 
surface brightness.  In Figure 10, models with $S_0 = 10^2$ 
erg g$^{-1}$ produce a bright ring with a dark shadow outside it.
As the ring and shadow propagate out through the disk, they
fade relative to the surface brightness of surrounding disk 
material. In Figure 11, models with $S_0 = 10^4$ erg g$^{-1}$ 
produce 
a bright ring that propagates out through the disk. The ring
is brightest where planets with radii of $\sim$ 1000--3000 km
form. Inside the ring, planet formation has saturated with
the formation of 3000 km objects. Outside the ring, large
planets have yet to form.

In the animations of Figures 10--11 included in the electronic 
version of this paper,  planet formation appears as a succession 
of waves flowing outward in the disk \citep[see also][]{kb02b}.
Slow growth from 1 km to $\sim$ 100 km produces little dust and
concentrates more and more material into objects with a smaller
geometric cross-section per unit mass. Thus, a
dark wave moves out through the disk. Dust formed during runaway 
growth lies in bright rings which appear as a bright wave 
moving out through the disk.  Once planet formation is complete,
a dark wave -- which heralds the disappearance of dust -- 
propagates out through the disk.  During this phase, the largest
bodies in the disk may coalesce to form larger planets.

\subsection{Planet Formation After a Stellar Flyby}

Planetesimal calculations demonstrate that embedded planets with 
$r_i \gtrsim$ 1000 km can produce bright rings and dark gaps in a 
quiet planetesimal disk. Stellar flybys can also produce rings, 
gaps, and possibly other structures in a planetesimal disk
\citep{lar97,mou97,kal00,ida00,kob01,lar01}. 
The lifetimes for rings and gaps produced by a flyby are much 
shorter, $\lesssim$ 1 Myr, than the lifetimes of rings
and gaps produced by planets, $\sim$ 10--100 Myr. Because field stars 
are less likely to interact with passing stars than stars in young
associations, flybys are more relevant to understanding the structures 
observed in young debris disks than in old debris disks
\citep[e.g.][]{ida00,kal00,kb02a}. Here we consider whether 
planet formation can regenerate bright rings and gaps following 
a stellar flyby.

Because our code does not currently follow the trajectories of 
individual objects, we assume that the close passage of a low 
mass star at $a \gtrsim$ 600 AU instantaneously raises the 
eccentricities of planetesimals. This approximation is valid if
the encounter between the passing star and the disk is short 
compared to the collisional damping time.  Passing stars not 
bound to the central star of the disk satisfy this requirement 
\citep{lar97,mou97}. We adopt a functional form for the imposed
eccentricity that satisfies constraints derived from $n$-body 
calculations \citep{lar97,mou97,kob01}.

\citet{kb02a} describe the early evolution of a planetesimal disk 
following a moderately close stellar flyby.  Large perturbations 
with $e_0 \gtrsim$ 0.05 lead to complete disruption of nearly all 
planetesimals \citep[][and references therein]{kb02a}.  When a flyby 
produces a modest perturbation with $e_0 \simless$ 0.03--0.04 at 
30--150 AU, two body collisions produce substantial amounts of dust 
and damp planetesimal velocities on short timescales.  Once 
planetesimals have $e \lesssim$ 0.01, collisions produce growth 
instead of disruption.  Continued damping and growth eventually 
produces planets.

To illustrate planet formation in a planetesimal disk following a
stellar flyby, we consider in detail a disk model with 
$e_0$ = 0.02 $(a_i/a_0)^{1/2}$,
$\Sigma_0$ = 0.02 g cm$^{-2}$ at $a_0$ = 30 AU,
and a total mass in solids of 9.3 $M_{\oplus}$.  This disk mass 
is at the lower end of the observed range of dust masses for 
0.5--3 $M_{\oplus}$ stars \citep{wya03a}. The disk consists of 
64 annuli with $\Delta a_i/a_i$ = 0.025 and extends from 30 AU 
to 150 AU.  The initial distribution of planetesimals has sizes of 
10 cm to 10 m with $\delta = 2$ and equal mass per mass bin. The 
particles have mass density $\rho_d$ = 1.5 g cm$^{-3}$.  We adopt 
the \cite{kb01} stirring algorithm and the \citet{wet93} 
fragmentation algorithm with $Q_c = 5 \times 10^7$ erg g$^{-1}$ 
and $S_0 = 2 \times 10^6$ erg g$^{-1}$. \citet{kb02a} describe 
several aspects of the early evolution of this model.

Planet growth following a stellar flyby has four stages.  After
the flyby, planetesimals damp quickly and begin to form larger 
objects.  This evolution starts at the inner edge of the disk
and slowly propagates outward.  At all heliocentric distances, 
the 50--200 cm planetesimals damp first.  This damping produce a 
V-shaped eccentricity distribution in $\lesssim$ 1 Myr
at 30--37 AU (Figure 12).  After 2--4 Myr, larger planetesimals 
damp and begin to grow slowly.  As objects grow to sizes of
0.1--1 km, dynamical friction dominates viscous stirring and 
reduces the eccentricities of the largest objects.  

When particles reach sizes of 1 km or larger, the growth and
stirring rates increase rapidly (Figure 13). Damping of the
eccentricities of the largest objects leads to runaway growth.
At 30--37 AU, it takes 40 Myr for the largest objects to reach 
sizes of 10 km and another 20 Myr for large objects to reach sizes 
of 100 km.  These timescales are a factor of $\sim$ 2 larger 
than growth times for icy planets in a quiet disk. 
Viscous stirring by the largest bodies increases the 
eccentricities of the smallest objects.  The velocity evolution
quickly reaches an equilibrium between the largest bodies with
radii of 100 km or larger and 1--10 km bodies containing most of 
the mass. This ratio is roughly $e_{large}/e_{small} \sim$ 
0.03--0.1 \citep{gol02}.  As the largest bodies grow to sizes 
of 1000 km, they stir up the smallest bodies to this equilibrium 
eccentricity ratio.

All processes take longer farther out in the disk.  The timescales
for collisions, damping, and stirring grow with heliocentric distance.  
The collision time is $t \propto P / \Sigma$ $\propto a^{3}$, which
implies a damping timescale of $\sim$ 0.3 Myr at 30--37 AU and 
$\sim$ 25 Myr at 130--150 AU. Because our model has 
$e_0 \propto a_i^{1/2}$, the outer disk damps more slowly and 
loses more material than the inner disk. This extra mass loss slows 
the rate of planet formation by another 20\% to 30\% relative to 
the inner disk.  Thus, planet formation takes $\sim$ 200 times 
longer at 150 AU than at 30 AU.

Throughout the evolution, the cumulative size distribution has three 
main components (Figure 14).  As collisions damp particle eccentricities,
fragmentation dominates growth by mergers.  The size distribution
for the small objects follows a power law with $\alpha_s \approx$ 2.5
for $r \lesssim$ 0.1--1 km.
Once mergers become important, the size distribution develops 
a merger component, with $\alpha_l \approx$ 3 for $r \gtrsim$ 
1--10 km. At intermediate
sizes, 0.1 km $\lesssim r \lesssim$ 10 km, the size distribution
has a pronounced hump containing most of the mass.  As planets
grow in the disk, the position of this hump moves to larger radii.

The right panel of Figure 13 illustrates some features of the 
collisional cascade
induced by planet formation.  During the early evolution of this
model, erosive collisions remove material from the disk.  The 
cumulative surface density declines by $\sim$ 30\% at 30--37 AU 
and 40\% to 50\% at 125---150 AU.  As mergers produce 1--100 km
objects, the cumulative surface density is roughly constant in time.
Once 100--1000 km objects begin to form, viscous stirring increases 
the eccentricities of the leftover planetesimals (Figure 12). 
Continued stirring leads to the collisional cascade, where 0.1--1 km 
planetesimals are slowly ground into smaller and smaller objects. 
Because 1--10 km planetesimals are too strong, the hump in the
size distribution shifts from $\sim$ 1 km to $\sim$ 10 km.
After 2 Gyr, the collisional cascade produces a factor of two
decline in the cumulative surface density.

Planet growth and the collisional cascade depend on the initial
mass in planetesimals.  More massive disks have shorter collision
times, with $t_c \propto M_d^{-1}$ \citep[e.g.,][1999]{lis87,kl98}. 
In a disk 
with a mass in solids comparable to the minimum mass solar nebula,
planets grow a factor of ten faster than in the model in Figures 11--13.
Figure 14 shows that the character of the evolution is not sensitive
to the initial mass.  The cumulative number distribution rapidly 
develops a smooth power law for $r \lesssim$ 0.1 km, with 
$\alpha_s \approx$ 2.5.
For $r \gtrsim$ 10 km, mergers lead to a steeper power law 
with $\alpha_l \approx$ 3 on longer timescales.  After 
$\sim$ 30 Myr, the largest objects have $r \sim$ 1000 km 
at 30--37 AU.  The collisional cascade begins and reduces 
the surface density by a factor of 3 over the next 300 Myr.
After 3 Gyr, erosive collisions reduce the surface density
at the inner edge of the disk by another factor of three.  The
largest objects then have radii of $\sim$ 2000--3000 km; most
of the mass is in objects with $r \sim$ 10--100 km.

To test the robustness of these results, we calculated models with a 
variety of initial conditions.  We changed $S_0$, $e_0$, and 
the initial variation of $e$ with $a$.  Besides the initial mass 
in planetesimals, the ratio 
$e_0/S_0$ is the important parameter in these calculations.  Compared 
to our baseline models with $e_0/S_0 = 10^{-8}$, models with 
$e_0/S_0 > 10^{-8}$ lose a larger fraction of their initial mass and
take longer to form planets.  When $e_0/S_0 \gtrsim 3 \times 10^{-8}$,
erosive collisions remove almost all planetesimals before collisional
damping becomes effective.  These calculations do not form planets
on reasonable timescales, $\lesssim$ 1 Gyr.  For calculations with
fixed $e_0/S_0$, the mass of the largest planet scales with $S_0$.  In 
our models, collisional cascades begin sooner when $S_0$ is smaller.
Because the collisional cascade robs material from the largest bodies,
smaller $S_0$ prevents large objects from growing.

Calculations with the \citet{dav85} fragmentation algorithm yield 
similar results.  Our tests show that damping is more efficient 
in calculations where $f_{KE}$ is large and less efficient when 
$f_{KE}$ is small.  In general, icy planetesimals with 
$f_{KE} \approx$ 0.03--0.10 are harder to disrupt and easier 
to fragment in the \citet{dav85} algorithm.  Thus, these models 
tend to produce less debris and to form planets on shorter
timescales than our baseline models.   This difference is small,
$\sim$ 25\% to 35\%.

Stellar flybys and planet formation produce copious amounts of dust.  
Figure 16 compares the evolution of the debris production rate
$\dot{M}$ of flyby models with the $\dot{M}$ evolution for a quiet
disk model.  During the flyby, collisions produce substantial debris
on a short timescale.  The derived $\dot{M}$ is 6--7 orders of
magnitude larger than the initial $\dot{M}$ for a quiescent disk
and 4--5 orders of magnitude larger than the largest $\dot{M}$ for 
a planet-forming disk.  Collisions damp planetesimal velocities;
$\dot{M}$ declines. As collisions start to favor mergers over
debris production, the decline in $\dot{M}$ accelerates. When
10 km objects begin to form in the inner disk, the global $\dot{M}$
reaches a minimum. The growth of 100 km objects leads to a linear
increase in $\dot{M}$; the growth in $\dot{M}$ accelerates once
1000 km objects begin to form and stir up the leftover planetesimals. 
Debris production saturates during oligarchic growth and then
declines as planet formation propagates out through the disk.

Compared to quiet disk models, planet formation following a 
stellar flyby produces less dust at later times (Figure 16). 
After a stellar flyby perturbs a planetesimal disk, planets grow
more slowly; once planet formation begins, the reservoir of
leftover planetesimals is smaller.  For a modest perturbation
with $e_0$ = 0.02, dust production following a flyby begins 
3--10 times later and peaks at a factor of two smaller $\dot{M}$ 
than dust production in a quiet disk.  Because A-type stars 
evolve into red giants on 1 Gyr timescales, stronger perturbations
probably prevent planet formation at 30--150 AU around A-type stars.

We conclude this section with a brief discussion of Figure 17,
which illustrates the evolution of the dust luminosity following
a stellar flyby.  The initial perturbation produces a substantial
dust luminosity that is 2--4 times larger than the maximum luminosity
of a planet-forming disk (compare with Figure 8). This large
luminosity is short-lived and declines rapidly as collisions 
cool the planetesimal swarm.  As objects grow to 10 km sizes in 
the inner disk, the dust luminosity reaches a minimum roughly
3 times more luminous than the luminosity minimum for a quiescent
planet-forming disk.  Dust production associated with planet 
formation leads to a rise in luminosity followed by a linear 
decline as in the quiet disk models. The peak luminosity in 
flyby models is significantly smaller than in quiet disk models. 
For similar initial masses, flyby models with $e_0$ = 0.02 are 
a factor of 3--5 fainter than quiet disk models. The flyby also 
causes a substantial delay in the rise in dust luminosity. Flyby 
models with $e_0$ = 0.02 reach maximum dust luminosity 5--10 
times later in time than quiet disk models.

\subsection{Limitations of the Models}

In previous papers, we have described limitations to multiannulus
\citep[][2002a]{kb01} and single annulus \citep[][1999]{kl98}
coagulation calculations.  Here, we summarize the most important of 
these limitations and consider uncertainties in the dust calculation.

As long as the statistical assumptions underlying the formalism are
met, coagulation calculations provide a reasonable representation of 
real collision evolution
\citep{wet80,gre84,dav85,bar91,spa91,lis93,wet93,st97,wei97,kl98,ina01}.
In our calculations, the spacing of mass bins in an annulus and the 
spacing of annuli in the disk limit the accuracy of the results. Our
choice of mass spacing, $\delta = 2$, lengthens the evolution time by
10\% to 20\% \citep[see][and references therein]{kl98}. The radial
resolution in our grid of annuli, $\Delta a_i/a_i$ = 0.025 also lengthens
the evolution time. Tests with different grids suggest a lag of 
15\% to 25\% relative to calculations with very fine grids. Thus,
our evolution timescales are $\sim$ 30\% to 40\% longer than the actual 
evolution times.

If $N$ is the number of annuli and $M$ is the number of mass batches,
the computation times increase roughly as $N^2 M^2$. Because a typical
calculation requires 2--4 weeks of cpu time, increasing the resolution
of a calculation to achieve a more accurate evolution time is 
prohibitively expensive.  Improving the accuracy of our evolution 
times requires practical increases in computing speed, which will 
be achieved with the next generation of high-speed parallel computers.

The coagulation algorithm begins to break down when binary interactions
between large objects become important. We reach this limit when
$r_i \approx$ 1000--2000 km. At this point, the coagulation algorithm
underestimates collision and stirring times, and overestimates the 
evolution time for dust clearing and the growth of the largest objects.
Tests with a hybrid $n$-body--coagulation code \citep{bk04} suggest
that our pure coagulation results overestimate the dust clearing time
by less than a factor of two. Although the hybrid code forms larger
objects than the pure coagulation code, differences in the final
mass distribution are small. We plan to report on these calculations 
in a future paper.

In our implementation, the inherent limitations of the coagulation
algorithm have clear observational consequences. We overestimate
the timescale to produce large planets by $\sim$ 40\% and the 
timescale for dust clearing by a factor of $\sim$ 2. Because the
collisional cascade must remove the same amount of material on a 
shorter timescale, our estimates for dust mass and luminosity are
probably low by a factor of $\sim$ 1.5--2.0 compared to ideal calculations
with perfect resolution in mass and radius. Our bright dust rings 
are probably 50\% larger in radius than those in an ideal 
calculation; our dark gaps are probably smaller by a similar 
factor. If the actual
clearing time for dust is shorter, then our radial surface brightness
profiles are too shallow during the late stages of the evolution.
All of the uncertainties in timescales are small compared to the
1--2 order of magnitude range in evolution timescales set by the
range in initial disk mass \citep[e.g.][]{wya03a} and the bulk 
properties of planetesimals. Thus, the observational uncertainties 
are set more by unknown physics than by limits in the calculations.

Our estimates for the evolution of dust mass, luminosity, and 
surface brightness rely on several assumptions,
(i) large bodies do not accrete small grains,
(ii) small grains have the same scale height as 1 m objects, and
(iii) the size distribution of dust grains is fixed at
$N_c \propto r_i^{-5/2}$.  
Large bodies probably accrete $\sim$ 1--3\% of the total mass 
converted to dust, which has a negligible impact on the evolution 
of small or large bodies.  Stirring by large planets sets the 
scale height of the smallest objects in the planetesimal grid.
The amount of stirring does not depend on the mass of the
small objects.  Because gas drag is unimportant at late stages 
in our calculations, the scale height of small objects is fairly
independent of their mass (Figures 1, 4, and 8). We estimate 
a factor of $\sim$ 2 uncertainty in the scale height of the
small objects, which yields factor of two uncertainties in
the dust luminosity.  \citet{doh69} and \citet{wil94} show that
collisional cascades produce power law size distributions with
$\alpha_s \approx$ 2.5 \citep[see also][]{tan96}. The uncertainty 
in this exponent is
small, $\pm$0.05 or less for a large range in input parameters.
In our calculations, significant deviations from this power
law can occur at a boundary size, $r_b$, where collisions
produce complete disruption for $r_i \lesssim r_b$ and
modest fragmentation for $r_i \gtrsim r_b$. For the bulk
properties assumed in our calculations, $r_b > $ 1 m, the
minimum size of planetesimals in our main calculation. Thus
we expect $\alpha_s \approx$ 2.5 for small particles to a
high degree of accuracy. This result yields a small uncertainty,
$\lesssim$ 5\%, in the derived dust masses and luminosities.

\section{DISCUSSION AND SUMMARY}

Our calculations demonstrate that icy planet formation is the
inevitable outcome of coagulation in the outer regions of 
a planetesimal disk \citep[see also][]{lis87,ken02}.  
In a quiet disk,  icy planets with 
$r_i \approx$ 1000--3000 km form on a timescale
\begin{equation}
t_P \approx 15-20 ~ {\rm Myr} \left ( \frac{\Sigma_0}{\Sigma_{MMSN}} \right )^{-1} 
\left ( \frac{a}{\rm 30 ~ AU} \right )^3 ,
\end{equation}
where $\Sigma_{MMSN}$ is the surface density of the minimum
mass solar nebula. Planets form more slowly in a disk
perturbed by a stellar flyby. Large perturbations prevent
planet formation. In modest perturbations with $e_0 \approx$
0.02--0.04, planet formation timescales are a factor of
2--4 longer than timescales in a quiet disk.

Icy planet formation produces copious amounts of dust. As planets
grow to sizes of 100--1000 km, gravitational stirring increases
the orbital eccentricities and inclinations of leftover planetesimals.
High velocity collisions between leftover planetesimals lead to
a collisional cascade, where 1~km objects are slowly ground to dust.
The collisional cascade reduces the surface density in the disk by
90\%--95\%. The reduction in surface density depends weakly on the
tensile strength $S_0$ of 1 km planetesimals. Collisions between
`weak' planetesimals with $S_0 \lesssim 10^3$ erg g$^{-1}$ produce
more dust than collisions between `strong' planetesimals with
$S_0 \gtrsim 10^3$ erg g$^{-1}$.

In all of our calculations, dust is confined to bright rings separated
by dark gaps. Bright rings form along the orbits of large planets,
which stir leftover planetesimals up to the disruption velocity. Dark
gaps are regions with little dust, where planets have not yet formed,
or regions shadowed by bright, optically thick dusty rings in the 
inner disk. The
bright rings and dark gaps have sizes, 0.1--0.2 $a$, comparable to
the rings and gaps observed in HR 4796A and other debris disks.
 
Dusty rings produced by icy planet formation are observable. The
maximum luminosity of a dusty disk, equation (2), is comparable
to the dust luminosities of known debris disk systems
\citep{bac93,kal98,lag00,hab01,spa01,gre03,dec03}. The decline in 
dust luminosity follows a power law, $L_d \propto t^{-m}$, 
with $m \approx$ 1--2 \citep[][see also equation (3)]{dom03}.
If the initial mass in planetesimals is comparable to the mass of
a minimum mass solar nebula, the timescale for the decline
in dust luminosity is 500 Myr to 2 Gyr. More massive disks evolve
on shorter timescales. The maximum dust luminosity agrees with
observed luminosities of debris disks. Our derived lifetimes are
longer than the observed lifetimes \citep[see also][]{gre03,dec03}. 
Limitations in our calculations probably lead us to overestimate
the decline timescales by a factor of $\sim$ 2--3.
Given the uncertainties in the initial conditions,
this `correction' brings our model lifetimes into
reasonable agreement with observations.

Aside from the initial disk mass and the mass density of
planetesimals, our results are insensitive to most input
parameters. The dust mass -- and thus the dust luminosity
and surface brightness -- does not depend on the bulk properties
or the initial orbital or mass distributions of the planetesimals.
The evolution timescales are longer if planetesimals are
more dense, stronger, and have larger initial orbital
eccentricities. Porous, weaker planetesimals in more
circular orbits grow more rapidly. For fixed initial
conditions, the uncertainties in the results are small
compared to the 1--2 order of magnitude range in the 
initial disk mass \citep{wya03a}.

Together with other studies, our results demonstrate that
several physical processes can produce rings, gaps, and
other structures in a debris disk. If the gas density
in the disk is modest, coupling between the gas and the
dust can produce rings and gaps \citep{tak01}. Resonant
interactions between dust and distant large planets
\citep{wil02,kuc03} or small moons \citep{wya99} can produce
rings or ring arcs in the disk.  Large planets can clear 
large gaps in the disk \citep{lin79,gol80}. Our models
produce bright rings from collisional cascades induced
by small planets within the ring. Dark gaps arise between 
small planets in nearby bright rings or by shadowing of 
the disk by an optically thick bright ring.

In all models, the visibility of dusty structures in a debris
disk is related to the collision rate. The collision lifetime 
of a single particle of radius $r$ in an annulus of width 
$\Delta a$ at distance $a$ from the central star is
$t_c \approx (dn/dt)^{-1}$:
\begin{equation}
t_c \approx \frac{a ~ \Delta a }{\Omega ~ n ~ r^2} ~ ,
\end{equation}
where $\Omega$ is the orbital frequency and $n$ is the number of
particles. The luminosity of the annulus is roughly 
$L_D/L_0 \approx \tau H / a \approx (r/a)^2 n$, which yields
\begin{equation}
t_c \sim 10 ~ {\rm Myr} \left ( \frac{\Delta a}{a} \right ) ~ \left ( \frac{P}{\rm 1000 ~ yr} \right ) ~ \left ( \frac{L_D/L_0}{10^{-5}} \right )^{-1} .
\end{equation}
Because the optical depth and collisional cross-section depend on
$r^2$, the collision time is independent of particle size.  Most 
debris disks have relative luminosities $L_D/L_0 > 10^{-5}$, 
orbital periods $P <$ 1000 yr, and relative sizes 
$\delta a /a \approx$ 0.1--0.2. The collision time of 10 Myr or less 
is shorter than the typical stellar age, $t \gtrsim$ 10--100 Myr. 
Particle collisions thus play an important role in the evolution 
of all known debris disks. Collisions are most important in the 
most luminous debris disks such as $\beta$ Pic and HR 4796A.

As samples of debris disk systems grow, observations might measure 
the importance of collisions.  In bright disks, very small grains 
ejected by radiation pressure produce a large optical depth.  
Systematic variations in the radial scale height with disk luminosity
or stellar age might imply some emission from these small grains.
Measured disk luminosities inside bright rings might constrain 
the amount of material dragged inwards by Poynting-Robertson drag
and thus the mass in small grains inside the bright ring. As our
calculations become more sophisticated, we plan to make detailed 
predictions for comparisons with observations.

Our results suggest caution when interpreting bright and dark 
structures in debris disks \cite[see also][]{tak01}. Dark gaps can
indicate a large planet \citep{lin79,gol80}, with a mass
$m_p \approx 3 (\Delta a/a)^3 M_{\star}$. For the measured sizes 
of dark gaps in known debris disks, $m_p/M_{\star} \simless 10^{-3}$.  
In our models, 
shadowing or several bright rings can produce dark gaps with 
$\Delta a /a \approx$ 0.01-0.1. These structures imply smaller 
planets with $m_p/M_{\star} \simless 10^{-7}$. In systems with 
residual gas, interactions between the gas and small dust grains 
can produce a single bright ring \citep{tak01} without any planet.

Observations can plausibly distinguish between these possibilities.
Multiple rings, apparently observed in $\beta$ Pic \citep{wah03}, 
favor planet formation over dust-gas interactions. 
Measured scale heights or velocity dispersions might distinguish
between large or small planets associated with a bright ring or 
a dark gap.  Large scale heights with $H/a \approx$ 0.1 are more 
consistent with Jovian mass planets; small scale heights with
$H/a \approx 10^{-2}$ imply smaller planets with masses ranging
from Pluto to the Earth.

Finally, multiannulus accretion codes are an important step towards 
understanding the formation and evolution of planetary systems
\citep[see also][]{spa91,wei97}.  Our calculations demonstrate that 
`complete' disk models with 64 or more annuli provide interesting
conclusions regarding icy planet formation in a planetesimal disk.
As improvements in computers allow larger simulations, we plan
to extend these calculations into the terrestrial and giant planet
regimes to improve constraints on the planetesimal theory. These
calculations also make interesting predictions regarding outcomes
of planet formation. Observations with satellites -- e.g.. 
{\it SIRTF} and {\it JWST} -- and ground-based telescopes -- 
e.g., {\it OWL} and {\it GSMT} -- will provide tests of these
predictions and lead to improvements in our understanding of
planet formation.

\vskip 4ex

We acknowledge a generous allotment, $\sim$ 1500 cpu days, of 
computer time on the Silicon Graphics Origin-2000 computers
`Alhena', `Castor', and `Pollux' at the Jet Propulsion laboratory
through funding from the NASA Offices of Mission to Planet 
Earth, Aeronautics, and Space Science.  Advice and comments 
from M. Geller, J. Greaves, and J. Wood greatly improved our 
presentation.  We thank P. Goldreich for clarifying our 
understanding of gravitational stirring in the low velocity
limit and for suggesting plots of cumulative surface density.
Portions of this project were supported by the {\it NASA }
{\it Astrophysics Theory Program,} through grant NAG5-13278.

\vfill
\eject

\clearpage

\appendix

\section{APPENDIX}

Our planetesimal calculations derive the evolution of the 
numbers and orbital elements for objects with radii of 1 m
and larger as a function of time \citep[2002a]{kb01}.
Detailed calculations for the evolution of the smaller 
particles is prohibitively expensive in computer time. 
To estimate the number of smaller particles, we perform 
a second, approximate calculation that uses the production 
rate $\dot{M}$ of small particles derived in the planetesimal
calculation\footnote{\citet{the03} perform a similar 
calculation in a single annulus and apply their results
to the inner disk of $\beta$ Pic.}.  Throughout the later 
stages of the planetesimal
calculations, $h_{ik} \approx$ constant for 1 m to 1 km objects. 
We thus assume that the scale height of small particles is 
identical to the scale height of 1 m objects in the planetesimal 
grid.  During a collisional cascade, \cite{doh69} and 
\cite{wil94} show that the cumulative number distribution of 
small objects is a power law, $N_C = N_0 r^{-\alpha}$, with 
$\alpha \approx$ 2.5.  At late times, the 1 m to 1 km objects 
in our planetesimal calculations reproduce this reult. We 
therefore assume that the dusty debris produced in the 
planetesimal calculations has the same power law.  The dust 
production rate in each annulus $\dot{M}_i$ thus yields the 
normalization constant $N_0$ of each annulus for each timestep.

We consider two populations of dust, very small grains and larger 
grains. On a dynamical time scale set by the local circular velocity,
$v_{K} = (G M / a)^{1/2}$, radiation pressure ejects very
small grains with radii between $r_1$ and $r_2$. If $\dot{M_k}_s$ 
is the production rate of very small grains in annulus $k$, the space 
density of very small grains ejected from annulus $k$ is
\begin{equation}
\rho_k = \frac{\dot{M_k}_s}{4 \pi a_k^2 v_{Kk} ~ {\rm sin}~\theta_k} ~ ,
\end{equation}
where $\theta_k$ is the opening angle defined by the vertical scale 
height of the very small grain population. The volume density $\rho_k$ 
depends on the size distribution of very small grains,
\begin{equation}
\rho_k = \frac{4 \pi \rho_g}{3} \int_{r_1}^{r_2} n_{0k} r^{3-\alpha}  ~ ,
\end{equation}
where $\rho_g$ is the mass density of an individual grain. We adopt 
$\alpha = 2.5$ and assume that grains have the same mass density as 
particles in the planetesimal grid.  Solving the two equations 
for $\rho_k$ yields the set of normalization constants $n_{0k}$.  
The normalization constants yield the volume density of the 
outflowing wind of very small grains as the sum of material 
ejected from all interior annuli
\begin{equation}
\rho_i = \sum_{k=1}^{k=i} \rho_k
\end{equation}

To derive the optical depth of the very small grains, we adopt the 
geometric optics limit 
\begin{equation}
\tau_{si} = 2 \pi \int \int n(r,a) a^2 da dr,
\end{equation}
where $n = n_{0i} r^{-\alpha}$ is the space number density.
We solve the integral over the size distribution exactly and
sum the optical depth through the planetesimal grid:
\begin{equation}
\tau_{s} = \frac{3 ( \sqrt{r_2/r_1} - 1 )}{8 \pi \rho_g r_2 ( 1 - \sqrt{r_1/r_2} ) } \sum_{i=1}^{N} ~ \left [ \sum_{k=1}^{i} \left ( \frac{\dot{M}_k}{v_{Kk} h_k } \right ) \left ( \frac{1}{a_{b,k}} - \frac{1}{a_{b,k+1}} \right ) \right ]~ ,
\end{equation}
where $a_{b,k}$ is the inner boundary of an annulus centered
at $a_k$.  

Although the exact expression for the optical depth is a
complicated function of grid variables, we derive a simple
expression based on characteristic quantities averaged over
the grid. For a reasonable grain population, $r_1$ = 0.01 $\mu$m 
and $r_2$ = 1 $\mu$m.  The optical depth is
\begin{equation}
\tau_s \approx 0.056 \left ( \frac{\dot{M}}{10^{21}~{\rm g ~ yr^{-1}}} \right )
\left ( \frac{30 ~ {\rm AU}}{a} \right )
\left ( \frac{10 ~ {\rm km ~ s^{-1} }}{v_K} \right )
\left ( \frac{10^{-2}}{h} \right )
\left ( \frac{1.5 ~ {\rm g ~ cm^{-3} }}{\rho_g} \right )
\left ( \frac{10^{-4} ~ {\rm cm}}{r_1} \right )
\end{equation}
When the production rate of very small grains exceeds 
$10^{22}$ g yr$^{-1}$, the optical depth approaches unity. 
This rate corresponds to the ejection of 1--2 Earth masses 
every million years.

To derive the time evolution of the number density and
optical depth for the larger grains, we consider three 
processes.  Erosive collisions between particles in the
planetesimal grid yield large grains with a size distribution
$N_C = N_0 r^{-2.5}$. Erosive collisions between these
particles remove material from the large grain population
but maintain the same power law size distribution. 
Poynting-Robertson drag preferentially removes small
grains and changes the size distribution. In each annulus,
we begin with a set of discrete mass bins with minimum 
radius $r_2$ and maximum radius $r_{max}$; $r_{max}$ is 
the radius of the smallest particle in the planetesimal 
grid. In most of our calculations, $r_{max}$ = 1 m.
At $t$ = 0, all of the large grain bins are empty.  For 
each timestep, 
we add particles to the bins using $N_C$ derived from the
dust production rates $\dot{M}_i$.  We compute the  
collision rates and outcomes for the largest mass bin in 
each annulus and scale this rate to derive collision rates 
and outcomes for smaller dust grains. We add or remove mass 
from each bin based on these rates. Finally, the change in
particle number due to Poynting-Robertson drag is 
\begin{equation}
\frac{dn_{ik}}{dt} = \lambda_{ijkl} n_{ik}
\end{equation}
where $\lambda_{ijkl}$ depends on the size, density,
and radiative properties of the grains, the stellar flux,
and the relative numbers of grains in adjacent annuli
\citep[e.g.][]{bur79}.
Because these properties change slowly with time, we
assume $\lambda_{ijkl}$ is constant during each timestep
and integrate $dn_{ik}/dt$ exactly to derive the amount
of material dragged through the grid.  This algorithm
yields the number of grains in each mass bin in each
annulus as a function of time.  To derive the optical 
depth of the dust in each annulus, we assume the geometric 
optics limit and sum the optical depth of each mass bin.

Our simple dust collision algorithm yields the optical 
depths in planetesimals and planets, large dust grains, and 
very small dust grains in a discrete set of concentric annuli
surrounding a star. The optical depth of grains dragged 
out of the innermost annulus by Poynting-Robertson drag 
is small compared to the optical depth in all of the
annuli; we ignore this contribution to the optical depth.  
Tests of the algorithm indicate that the optical depths
are accurate to a factor of 1.5--2.0, which is adequate
for our purposes.

To estimate the emergent luminosity from the disk, we
assume that the central star is the only radiation source.
We follow \cite{kh87} and assume a spherical, limb-darkened 
star with radius $R_0$, luminosity $L_0$, and 
limb-darkening coefficient $\epsilon_0$ = 0.6. For a point
$P$ at the outer boundary of annulus $i$ with height $h_P$
above the disk midplane, rays from the star enter the annulus 
at a scale height $h_{in}$ above (below) the midplane.
We compute the length $l$ of the path through the disk and 
derive the optical depth along this path as 
$\tau_p$ = ($l/\Delta a_i$)$\tau_i$, 
where $\Delta a_i$ is the width of the annulus.  The radiation
absorbed along this path is $e^{-\tau_p} I_0$, where $I_0$ is
the flux incident on the boundary of the annulus. Numerical
integrations over the stellar surface and the vertical extent
of an annulus yield the amount of flux absorbed by each annulus,
which we convert to relative surface brightness. A final
numerical integration over the radial extent of the disk yields 
the ratio of the disk luminosity to the stellar luminosity,
$L_D/L_0$.

Although this algorithm is an improvement over our previous
optically thin treatment of radiative transfer through a
planetesimal disk, it has several limitations.  The optical
depth calculation does not distinguish between absorption
and scattering.  For a grain albedo $\omega$, the luminosity
in scattered light is roughly $\omega L_D/L_0$; the
thermal luminosity is roughly $( 1 - \omega ) L_D/L_0$.
The algorithm also ignores multiple scattering and absorption.
Because the vertical optical depth of the disk is $\sim 10^{-3}$ 
or smaller in most cases, this approximation is satisfactory.

\clearpage

\centerline{\bf ANIMATIONS and FIGURES}
\vskip 4ex

\noindent
To download the figures for this paper, including two
color stills, go to

\vskip 4ex

\noindent{http://cfa-www.harvard.edu/$\sim$kenyon/pf/emb-planet-figures.pdf}

\vskip 4ex

\noindent
To view the two animations described in the paper, go to

\vskip 4ex 
 
\noindent{http://cfa-www.harvard.edu/$\sim$kenyon/pf/emb-planet-movies.html}

\end{document}